\documentstyle[rotate,amssymb,psfig,eqsecnum,aps]{revtex}

\input{epsf}

\renewcommand{\arraystretch}{1.3}

\begin{document}
\title{Observation of Anomalous Dimuon Events in the NuTeV Decay
Detector\\
(Preliminary)\\
}
\author{T.~Adams$^{4}$, A.~Alton$^{4}$, S.~Avvakumov$^{8}$, 
L.~de~Barbaro$^{5}$, P.~de~Barbaro$^{8}$, R.~H.~Bernstein$^{3}$, 
A.~Bodek$^{8}$, T.~Bolton$^{4}$, J.~Brau$^{6}$, D.~Buchholz$^{5}$, 
H.~Budd$^{8}$, L.~Bugel$^{3}$, J.~Conrad$^{2}$, R.~B.~Drucker$^{6}$, 
B.~T.~Fleming$^{2}$, R.~Frey$^{6}$, J.~Formaggio$^{2}$, J.~Goldman$^{4}$, 
M.~Goncharov$^{4}$, D.~A.~Harris$^{8}$, R.~A.~Johnson$^{1}$,
J.~H.~Kim$^{2}$, S.~Koutsoliotas$^{2}$, M.~J.~Lamm$^{3}$, W.~Marsh$^{3}$,
D.~Mason$^{6}$, J.~McDonald$^{7}$, C.~McNulty$^{2}$, K.~S.~McFarland$^{3}$, 
D.~Naples$^{7}$, P.~Nienaber$^{3}$, A.~Romosan$^{2}$, W.~K.~Sakumoto$^{8}$, 
H.~Schellman$^{5}$, M.~H.~Shaevitz$^{2}$, P.~Spentzouris$^{2}$, 
E.~G.~Stern$^{2}$, N.~Suwonjandee$^{1}$, M.~Vakili$^{1}$, A.~Vaitaitis$^{2}$, 
U.~K.~Yang$^{8}$, J.~Yu$^{3}$, G.~P.~Zeller$^{5}$, and E.~D.~Zimmerman$^{2}$}
\date{September 1, 2000}
\address{
$^1$University of Cincinnati, Cincinnati, OH 45221 \\
$^2$Columbia University, New York, NY 10027 \\
$^3$Fermi National Accelerator Laboratory, Batavia, IL 60510 \\
$^4$Kansas State University, Manhattan, KS 66506 \\
$^5$Northwestern University, Evanston, IL 60208 \\
$^6$University of Oregon, Eugene, OR 97403 \\
$^7$University of Pittsburgh, Pittsburgh, PA 15260 \\
$^8$University of Rochester, Rochester, NY 14627 \\
}

\maketitle

\begin{abstract}
A search for long-lived neutral particles ($N^0$) which decay into at
least one muon has been performed using an instrumented decay channel
at the E815 (NuTeV) experiment at Fermilab. The decay channel was composed
of helium bags interspersed with drift chambers, and was used in conjunction
with the NuTeV neutrino detector to search for $N^0$ decays. The data were
examined for particles decaying into the muonic final states $\mu \mu$, 
$\mu e$, and $\mu \pi$. Three $\mu \mu $ events were
observed over an expected background of $0.040 \pm 0.009$ events; no events
were observed in the other modes. Although the observed events share some 
characteristics with neutrino interactions, the observed rate 
is a factor of 75 greater than expected.  No Standard Model process 
appears to be consistent with this observation.

\vspace{0.1in} \noindent . 
\end{abstract}




\noindent
\parbox{\textwidth}{
 \parbox{8.6cm}{

\section{INTRODUCTION}

In various extensions to the Standard Model, new particles exist
which have reduced couplings to normal quarks and leptons. These new
particles may have zero electric charge, long
lifetimes, and small interaction rates with normal matter. We shall
refer to these as $N^0$ particles in the following text. 
Examples of such $N^0$ particles include neutral heavy leptons (NHLs) 
or heavy sterile neutrinos~\cite{grl,shrock,tim} and
neutral supersymmetric particles~\cite{liubo_susy} such as neutralinos
and sneutrinos. The $N^0$ particles can be produced either by pair
production in hadronic interactions or via weak decays of mesons
through mixing with standard neutrinos. The decays of the $N^0$ to
normal hadrons and/or leptons can proceed through weak decays with
mixing, or via $R$-parity violating supersymmetric processes.

\hspace{0.3cm}
High energy neutrino beamlines are ideal places to produce $N^0$ 
particles, since very large numbers of protons interact in these beamlines.  
$N^0$'s may be produced via a number of mechanisms, 
including primary interactions of the protons either in the target
or the beam dump, through prompt decays of charmed or bottom
mesons, by ~ decays ~of pions ~or kaons in the decay region, \hfill or in 
 } \hfill
 \parbox{8.6cm}{
neutrino interactions in the shielding downstream of the decay region.  
A particle detector placed downstream of this sort of beamline (i.e., in 
the neutrino beam itself) can be used to search for $N^0$ decays.

\hspace{0.3cm}
We report here the results of a search using Fermilab's E815 
(NuTeV) detector for $N^0$ particles in the mass region above 2.2 GeV/$c^2$ 
which decay into final states with at least one muon and one other 
charged particle. For the search described here, the NuTeV neutrino 
beamline was used in conjunction with a low mass decay detector 
called the decay channel.

\hspace{0.3cm}
NuTeV has previously reported results of searches for $N^{0}$'s in the
mass region between 0.3 to 3.0 GeV/$c^2$ with at least one final state
muon~\cite{arturv_prl}, and in the mass region below 0.3 GeV/$c^2$ for decays
to electrons~\cite{karmino_prl}. The 0.3 to 3.0 GeV/$c^2$ study addressed
NHLs that could be produced in the decay of K and D mesons, whose hadronic
production rate is known~\cite{charm}. This mass region also has low background
from deep inelastic neutrino events in the decay channel. The low mass
($<0.3$~GeV/$c^2$) study was pursued mainly to address the KARMEN timing
anomaly~\cite{karmen}, which has been interpreted as a $N^0$ particle
with a mass equal to 33.9 MeV/$c^2$.

\hspace{0.3cm}
The search for events with masses above 2.2 GeV/$c^2$ (which we shall refer
to as ``the high mass region'') is different from the previous
searches in two respects. First, 
 }
}

\twocolumn

\noindent
the backgrounds from neutrino 
interactions are much higher in the high mass range. Reducing this 
potential background source required tighter selection criteria than 
the ones used in the previous analyses.  Second, the production 
mechanisms are quite different from those for lower mass NHLs. 
NHLs with masses above 1.8 GeV/$c^2$ arise either from decays of B 
mesons or from neutrino interactions in the neutrino beam shielding 
(``berm'') upstream of the detector~\cite{albright}. The B meson 
production cross-section at 800 GeV/c is much smaller and 
less well known than that for the lighter mesons, and this partially
motivated separating the high mass search from the previous ones. 
Recently, though, new data from Fermilab E771 have indicated a larger 
B production cross-section, making a search for NHLs from this source 
feasible~\cite{bprod}.  

\section{THE BEAMLINE AND DETECTOR}

During the 1997 fixed-target run at Fermilab, NuTeV received
$2.54\times 10^{18}$ 800 GeV/c protons with the detector configured for
this search. The proton beam was incident on a one-interaction-length
beryllium oxide target at a targeting angle of 7.8 mr with respect to
the detector. A sign-selected quadrupole train (SSQT)~\cite{SSQT} focused
either positive (for $1.13\times 10^{18}$ protons) or negative (for
$1.41\times 10^{18}$ protons) secondary $\pi $ and $K$ mesons into a
440 m evacuated decay region pointed towards the NuTeV decay channel
and neutrino detector hall. Surviving neutrinos (and possibly also
$N^{0}$'s) traversed $\sim $850 meters of earth-berm shielding before
reaching the NuTeV decay channel.

\begin{figure}
 \unitlength1cm
 \begin{picture}(8,3.6)(0,0)
  \put(0,0){\psfig{figure=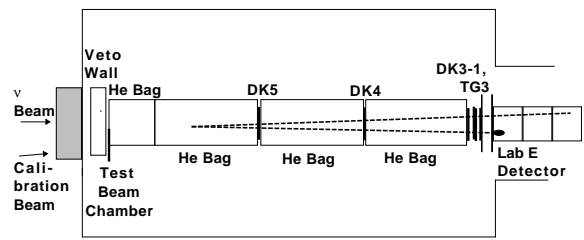,width=0.45\textwidth,clip=T}}
 \end{picture}
\caption{Schematic of the NuTeV decay channel with example $N^0 \rightarrow
\mu \pi$ decay.}
\label{fig:dkchannel}
\end{figure}

The decay channel region (Figure~\ref{fig:dkchannel}), located 1.4 km
downstream of the production target, was designed to contain minimal
material (in order to suppress neutrino interactions) and to have
tracking sufficient to isolate two-track decays of neutral
particles. A 4.6~m~$\times$~4.6~m double array of plastic scintillation
counters vetoed charged particles entering from upstream of the decay
channel.  If two counters in a back-front coincidence fired,  
the event was vetoed.  Timing resolution for the veto system was 3.8 ns.  
A NuTeV testbeam chamber was positioned immediately downstream of the veto 
wall and offset from the center of the decay channel; this chamber was 
not used in this analysis.  The channel itself measured 34 m in length 
and was interspersed with 3~m~$\times$~3~m argon-ethane drift chambers
positioned at 14.5~m, 24~m, and 34~m downstream of the veto array
in stations of 1, 1, and 4 chambers, respectively. From upstream to
downstream, the chambers were labeled DK5 through DK1, followed by
TG43 at the front face of the neutrino detector. Chambers DK5 through 
DK2 and TG43 had a single sense wire per cell (``single-wire'' chambers), 
resulting in a left-right ambiguity in the hit position.   This ambiguity 
was reduced in tracking by making use of the fact that the positions 
of the single-wire chambers were staggered in $x$ and $y$.  Chamber DK1 
was of a ``three-wire'' (two sense wires and one field-shaping wire 
per cell) design, which helped resolve any remaining tracking ambiguities.
DK1 was also rotated by 47.7 mr about the beam axis, allowing
tracks in the $x$ view to be matched with those in the $y$ view.
The regions between the drift chamber stations were occupied by helium-filled
cylindrical plastic bags 4.6~m in diameter.  

In the offline analysis, tracks were reconstructed from drift chamber 
hits and grouped together to form vertices. The tracking algorithm took 
into account multiple Coulomb scattering, using a full error matrix
for the fit.  Sets of tracks were grouped as candidates for a vertex if 
their distance of closest approach was less than 12.7 cm.   The vertex 
position was then determined using a constrained fit. Typically, a vertex
from a $N^0$ of mass 5 GeV/$c^2$ would be reconstructed with a resolution of
0.13 cm in the transverse direction and 7.4 cm longitudinally.

The Lab E neutrino detector~\cite{detector,nim}, located
immediately downstream of the decay channel, provided final state
particle energy measurement and identification. This detector
consisted of a 690 ton iron-scintillator target calorimeter followed
by a toroidal muon spectrometer. Three-wire argon-ethane drift
chambers were positioned every 20~cm along the length of the
calorimeter, and 84 2.5 cm-thick liquid scintillator counters were
interleaved with the steel plates at 10 cm intervals throughout the 
its length.   The spectrometer had a 15 kG toroidal 
magnetic field with drift chambers interspersed throughout the 
toroid magnets to provide tracking for the muons.

Sets of hits in the calorimeter drift chambers were linked to tracks found
within the decay channel to determine the particle identification 
for each track. By analyzing the distribution of hits in the calorimeter 
as either single, long tracks (consistent with a muon), as a compact 
cluster of hits (consistent with an electron shower), or as an 
elongated cluster of hits (consistent with a pion shower), the particle 
type was determined.  All charged hadrons were reconstructed as pions;
``$\mu\pi$'' samples referred to below are understood to include any
$\mu p$ or $\mu K$ final states as well.  The NuTeV calibration beam
provided electrons, pions, and muons to the Lab~E detector; 

\begin{table}
\caption{Misreconstruction rates for Monte Carlo $N^0$ decays \label{misid}}
{\center \begin{tabular}{cccc}
Generated & \multicolumn{3}{c}{Reconstruction mode probability} \\ \cline{2-4}
Decay Mode & $\mu \mu$ & $\mu e$ & $\mu \pi$ \\ \hline
$\mu \mu \nu$ &  $100^{+0.0}_{-0.3} \%$ & $0.0_{-0.0}^{+0.3}\%$ & $0.0_{-0.0}^{+0.3}\%$\\
$\mu e \nu$ &    $0.0_{-0.0}^{+0.3}\%$  & $89 \pm 2\%$ & $11 \pm 2\%$ \\
$\mu \pi $ &  $1 \pm 1 \%$ & $25 \pm 3 \%$ & $74 \pm 3 \%$  \\
\end{tabular}}
\end{table}

\bigskip
  
\noindent
these were used to tune the particle identification algorithm.
Misidentification rates for $N^0$ decays were determined using 
the GEANT-based Monte Carlo~(MC)~\cite{geant} described in 
Section~\ref{sec:bkgdest}; these rates are listed in 
Table~\ref{misid}. 

In the case of electrons and pions, pulse height information from 
the counters was used to determine particle energy deposition.  
The hadronic energy resolution of the calorimeter was 
$\sigma/E=(0.024\pm 0.001)\oplus(0.874\pm0.003)/\sqrt{E}$; 
the electromagnetic energy resolution,
${\sigma }/{E} =(0.04\pm 0.001)\oplus{(0.52\pm 0.01)}/{\sqrt{E}}$~\cite{nim}.
If two or more pion- or electron-clusters were present, the energy 
determined from the pulse heights was divided according to
the number of drift chamber hits in each cluster. 

Muon energy determination depended on the topology of the track.   
If the muon  track extended into the toroid, the spectrometer 
measurement was used (resolution of 11\%).   If the muon stopped 
in the calorimeter steel, the momentum was determined by range 
(resolution of 310 MeV). The energy of muons which exited through 
the side of the calorimeter was determined from the track's 
multiple scattering in the steel.  For a 50 GeV/$c$ muon, the 
resolution for this method is 42\%.

\section{PHILOSOPHY OF THIS ANALYSIS}

In analyses such as this, there is a real concern that events may be
eliminated or isolated through an unintentional bias of the people
involved in the analysis. The solution adopted by many
collaborations is that of a ``closed box'' analysis, in which 
there is no direct access to the signal region until the end of the analysis. 
This procedure was a philosophical goal of this search.  However, 
before this analysis, during the early development of the reconstruction 
software for the decay channel, one candidate decay channel event 
with two muon tracks was observed. This event was studied in detail 
and ascertained to have a mass greater than 2.2 GeV/c$^{2}$.

Because of the observation of this event, the NuTeV collaboration went to
considerable effort to minimize bias. Investigations of data events
with high mass were stopped until the MC background
studies (described below) were completed to establish the cuts and
requirements. In most cases, cuts set prior to the observation were
used. In those cases where new cuts were introduced, demonstration of a 
strong MC-based motivation was required. New members who
had not seen the event joined the analysis group. Finally, an important
aspect of the analysis included setting up orthogonal analysis regions and
comparing Monte Carlo prediction to the result. Each of the two 
previously-published analyses was motivated by its own physics goals, 
but they also represent tests in regions complementary to this analysis.

As part of the analysis philosophy, once the analysis region was
selected based on the Monte Carlo criteria, the collaboration agreed to
show any events which were observed. However, the interpretation of the
events might change after the analysis region was examined, upon further
investigation.

\section{EVENT SELECTION}

Event selection criteria were developed to minimize known backgrounds
while maintaining efficiency for a possible $N^0$ signal.   

\begin{figure}
\centerline{\psfig{figure=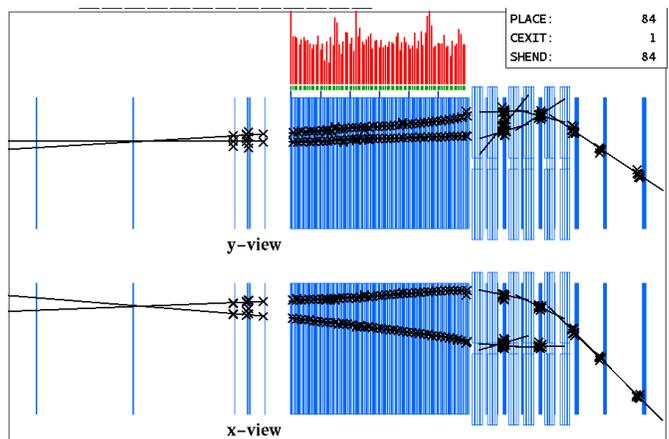,width=0.5\textwidth,clip=}}
\caption{Monte Carlo simulated event for a 5 GeV/$c^2$ $N^{0} 
\rightarrow \mu\mu\nu$.
\label{mumunu_evt}}
\end{figure}

As an example of what might be observed in the decay channel, 
Figure~\ref{mumunu_evt} shows a Monte Carlo simulated event of a 
5 GeV/$c^2$ $N^{0} \rightarrow \mu\mu\nu$ on the NuTeV event display.   
The beam enters the decay channel from the left of the figure.   
The decay channel chambers (DK5 through TG43) appear sequentially 
from left to right.  The vertex of the simulated decay is immediately 
downstream of DK4.   Hits in chambers are indicated by crosses; 
two hits per track are shown in each chamber because of the left-right 
ambiguity in the single-wire chambers.  The muons can be seen to
penetrate the calorimeter steel and then bend in the toroidal magnetic
field at the far right.  Pulse height information from the counters
is shown by the histogram above the calorimeter region.

\bigskip

\begin{minipage}{\textwidth}
\begin{center}
\begin{minipage}{14cm}
\begin{table*}
\caption{Overview of reconstruction cuts used in the analysis. }
\label{reconcut}
\begin{tabular}{c|c}

{{ Event conditions}} & No veto, physics trigger, 
protons in the spill, run period \\ \hline
{ Cut cosmic ray tracks} & slope $< 0.1$ radians \\ \hline
{{ 3-D track reconstruction}} & $x$ and $y$ views linked for 
each track \\ \hline
{{ Track \& vertex reconstruction}} & Track pseudo-$\chi^2$/dof$\le 10$ \\
& Vertex pseudo-$\chi^2$/dof$\le 10$ \\ \hline
{{ Require two-track vertex}} & Only two tracks connected to upstream vertex\\
 & only one particle downstream.\\ \hline
{{ Good particle ID}} & Track is within calorimeter acceptance\\
& Cluster associated with track \\
& $\mu$ Energy $> 2.2$ GeV, $e$ and $\pi$ energies $> 10$ GeV \\ \hline
{ {Cut $K_L$ punch-through}} & ${ \Sigma} E_{tracks} > 12$ GeV\\ \hline
{{ Isolate fiducial volume}} & vertex within $|x|<127, |y|<127$ cm \\
 \hspace{5.5cm} & vertex 101.6 cm away from chambers\\
& or 3 times vertex error (whichever is larger).\\ \hline
{{Isolate mass region}} & $M_T > 2.2$ GeV \\ 
\end{tabular}
\end{table*}
\end{minipage}
\end{center}
\end{minipage}

\bigskip

In decays with a neutral particle in the final state, it is not 
possible to reconstruct the invariant mass of the $N^0$.   Instead, 
one must use the ``transverse mass,'' 
$m_{T}\equiv|P_{T}|+\sqrt{P_{T}^{2}+m_{V}^{2}}$, where $P_{T}$ is the
component of the total reconstructed momentum perpendicular to the
beam direction, and $m_{V}$ is the invariant mass of the visible particles. 
In the case of perfect resolution, the transverse mass of the event 
is always lower than the $N^0$ mass.  When there is no final state 
neutrino, and hence no missing $P_T$, this expression reduces to 
the invariant mass. 

The main sources of conventional events in the decay channel are: 
1) deep-inelastic scattering (DIS) of neutrinos or anti-neutrinos 
in the drift chamber material;
2) DIS in the helium; and
3) DIS in the material surrounding the decay channel with a 
misreconstructed vertex in the fiducial volume.   
The DIS event rate rises with $M_T$, and represents the dominant 
background for the high mass analysis.

Other sources of background are small compared to DIS.   Neutral kaons 
produced in surrounding material may enter the channel and decay in 
the fiducial volume.  These will appear to have high transverse mass 
if the kaon enters with a large angle. Diffractive $\pi$, $K$, $\rho$, 
and charm production from neutrino interactions may occur in 
either the chambers or the helium.  Another interaction which 
can occur in either chambers or helium is low multiplicity 
neutrino-induced resonance production, characterized by a high-energy 
forward muon accompanied by a low-energy pion track.  A number of 
other possible background sources have been found to be negligible 
because \hfill they very rarely \hfill produce \hfill reconstructed 
\hfill vertices \hfill in the
\newpage

\vspace*{9.25cm}

\noindent
decay channel.   These include cosmic ray showers, 
conversions of photons produced in surrounding material, interactions
from muons scattered from surrounding 
material, ``leakage'' of charged particles from the adjacent 
testbeam line (which occurred only during specific 
data collection periods), an out-of-time neutrino interaction overlaid
on an in-time interaction, and two coincident in-time interactions in the 
decay channel.   These background sources were constrained by data 
as well as investigated through Monte Carlo.   

The goal of this stage of the analysis was to create a set of cuts 
which reduced the number of events expected from conventional 
sources to well below one event.  Most of these cuts were 
originally developed for the 0.3 to 3.0 GeV/$c^2$ search. The 
cuts used to isolate $N^0$ decays fell into two broad categories:
reconstruction and ``clean'' cuts.  

Reconstruction cuts isolated events with exactly two tracks forming 
a vertex within the decay channel fiducial volume and having no 
charged particle identified in the upstream veto system. A summary 
of these cuts appears in Table~\ref{reconcut}.  Both tracks were 
required to be well-reconstructed.   A 
``pseudo-$\chi_{trk}^{2}/{\rm dof}$'' (degrees of freedom) was 
used\footnote{``Pseudo'' means that Gaussian errors on hit positions 
were assumed.} 
to measure track reconstruction quality.   Tracks were required 
to have an associated calorimeter cluster, with at least one 
of the tracks identified as a muon. Cuts were also applied on 
vertex quality (pseudo-$\chi_{vx}^{2}/{\rm dof}<10$, which 
corresponds to a 95\% acceptance probability) and transverse 
position within the detector fiducial volume $(|x|<127$~cm,
$|y|<127$~cm).  In order to remove events which might be due to 
interactions in the chambers, events were cut where the longitudinal 
($z$) distance from the vertex position to any drift chamber was 
less than either $\pm$101.6 cm or three times the vertex error.
Events were allowed to have a second vertex downstream of the first to 
allow  for the possibility of events with $\delta$-rays.  Cosmic ray
tracks, which generally have large angles with respect to the beam direction,
were removed by requiring the slope of each track to be less than 100
mr. To ensure accurate particle identification and energy measurement,
muons, hadrons, and electrons were required to have an energy greater 
than 2.2~GeV, 10~GeV, and 10~GeV, respectively. 
This latter cut also eliminated low-energy pions associated
with neutrino-induced resonance production. An additional total
energy cut of 12 GeV was applied to $\mu \mu $ events to
remove background from $K_L \rightarrow \pi \mu \nu$ decays with a
subsequent decay of the pion.  In order to isolate high mass events, 
a cut of $m_{T}>2.2$~GeV/$c^2$ was applied.

``Clean event'' cuts were applied to reduce the deep-inelastic neutrino
scattering backgrounds.   These cuts, specific to this analysis, 
were motivated by the observation that DIS events typically have 
large track multiplicities with many drift chamber hits and extra, 
unassociated  clusters in the calorimeter. A summary of the 
characteristics associated with DIS events which pass the 
reconstruction cuts is given in Table~\ref{clncut}.  The ``clean'' cuts
included the following requirements: 1) three or fewer tracks in any
one view, 2) three or fewer drift chamber hits in any view of the
first chamber downstream of the vertex, 3) at least one view ($x$ or
$y$) with fewer than eight drift chamber hits total in the first
two chambers downstream of the vertex, 4) no energy clusters in the
calorimeter not associated with tracks, and 5) no tracks identified as
electrons with missing hits in either view of the first two
chambers downstream of the vertex. Requirements 1-3 remove events 
with high multiplicities; requirement 4 removes events where a 
neutral particle deposits energy in the calorimeter; \hfill and \hfill
requirement \hfill 5 

\begin{minipage}{\textwidth}
\begin{center}
\begin{minipage}{16cm}
\begin{table}
\caption{Identification of events with 
exiting tracks, neutral particles, and photon conversions; these
characterize DIS events and are removed by the ``clean'' cuts.}
\label{clncut}
\begin{tabular}{c|c}
Extra tracks  & $>$3 tracks in either view \\ \hline
Exiting tracks & $>$1 extra hit in either view in the chamber just downstream of vertex\\ 
\hspace{3cm} & or $>$7 hits in each view in the first 2 chambers downstream 
of the vertex.\\ \hline
Neutral particles & $\ge$ 1 cluster(s) in calorimeter without associated track.\\ \hline
Photon conversion &  Electron PID with no hits in chamber immediately 
downstream of vertex.\\ 
\end{tabular}
\end{table}
\end{minipage}
\end{center}
\end{minipage}

\newpage
\noindent
was used to reduce events with photon conversions 
in the downstream chambers which could be misidentified as electrons.

\section{BACKGROUND ESTIMATION USING MONTE CARLO} \label{sec:bkgdest}

Detailed Monte Carlo simulations of both physics processes and detector 
effects were used to quantify the background from neutrino interactions
after cuts.   Input to the simulation was provided from several 
event generators.  The LEPTO/Jetset Monte Carlo program was used to 
simulate DIS events~\cite{lund}.  This simulation used CCFR parton 
distributions~\cite{Bill}, included the correct A-dependence~\cite{EGS}, 
and generated DIS events from $Q^2 > 0.1$~GeV$^2$ and $W>2$~GeV. 
Resonance and continuum production, simulated using the calculations
of Belusevic and Rein~\cite{lownu}, allowed us to extend the
Monte Carlo into the low-$W$ region.  Diffractive production was 
calculated using Vector Meson Dominance (VMD) and 
Partially-Conserved Axial Current (PCAC) models normalized to a 
previous measurement with the NuTeV calorimeter data~\cite{diffnutev}. 

The event generators fed a GEANT-based~\cite{geant} detector 
simulation that produced hit-level simulations of raw data.  
Cell-by-cell inefficiencies and dead regions due to internal chamber 
supports were included.   To simulate noise and accidental activity 
in the detector, decay channel hits taken from in-time downstream
calorimeter neutrino events were overlaid on the GEANT events.
Monte Carlo events were processed using the same analysis routines
used for the data. 

Background calculations were normalized to the data using
charged-current DIS interactions in the chambers.  Events in this 
sample were required to pass the following five ``normalization cuts'': 
a vertex within the transverse fiducial volume 
($|x|<127$~cm, $|y|<127$~cm);  a $z$ vertex within 76.2 cm of a
drift chamber; no coincidences within $\pm 50$~ns of the trigger in 
the upstream veto system; $\ge$1~GeV energy deposit in the front 
of the calorimeter; and one toroid-analyzed muon matched to a 
decay channel track.  The Monte Carlo was normalized to \hfill match 

\clearpage

\begin{figure}
 \centerline{\psfig{figure=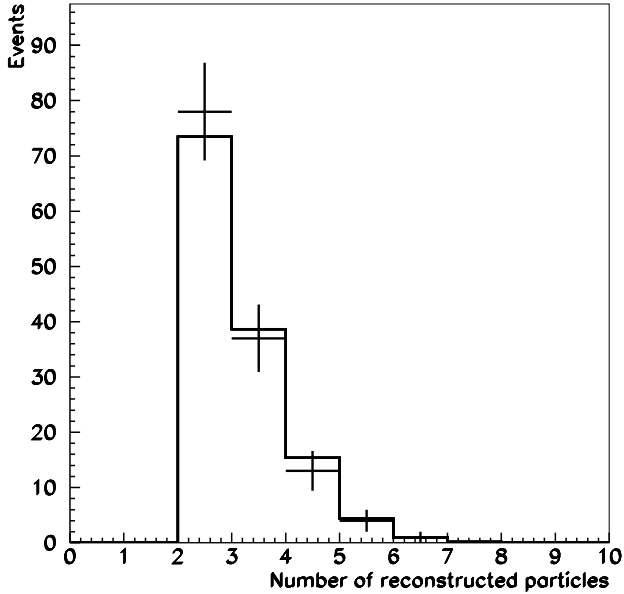,width=0.5\textwidth}}
 \caption{Distribution of number of reconstructed decay channel particles 
  comparing data (points) and Monte Carlo (histogram).  The samples
  are normalized to each other; this normalization is used for the
  final background estimates. \label{fig:ndkpart}}
\end{figure}

\bigskip

\noindent
the total number of data events with two or more tracks.  
Figure~\ref{fig:ndkpart} shows a comparison of the data and
MC distributions for this sample.  The preliminary error on this 
normalization is $9\%$.  Two alternate normalizations were used as 
checks.  The first normalized to protons on target using the decay 
channel mass distribution and (anti-)neutrino cross-sections.  The 
second normalized to charged-current interactions in the calorimeter, 
and scaled by the decay-channel-to-calorimeter mass ratio.  These 
cross-checks had uncertainties of 16$\%$ and 12$\%$ respectively, and 
were in agreement with the primary normalization.

Monte Carlo events were compared to data as a check on the quality of
the simulation. In such a comparison, the challenge is to isolate events 
of sufficiently similar topology to verify the Monte Carlo calculation 
of the background and at the same time maintain high statistics in 
the data sample.  We used two methods to achieve high statistics 
comparison samples of events with two or more tracks. 

For the first sample, the vertex was required to be within the decay
channel fiducial volume, with $|x|<127$~cm, $|y|<127$~cm, allowing
the $z$ position to be either in the chambers or the helium.   Tight
track angle cuts were imposed to remove cosmic rays, and a strict
requirement on veto system activity was used to remove upstream
interactions. Other cuts on reconstruction, particle
identification, and vertex fit quality were \hfill removed.  

\newpage

\vspace*{1.cm}

\begin{figure}
\centerline{\psfig{figure=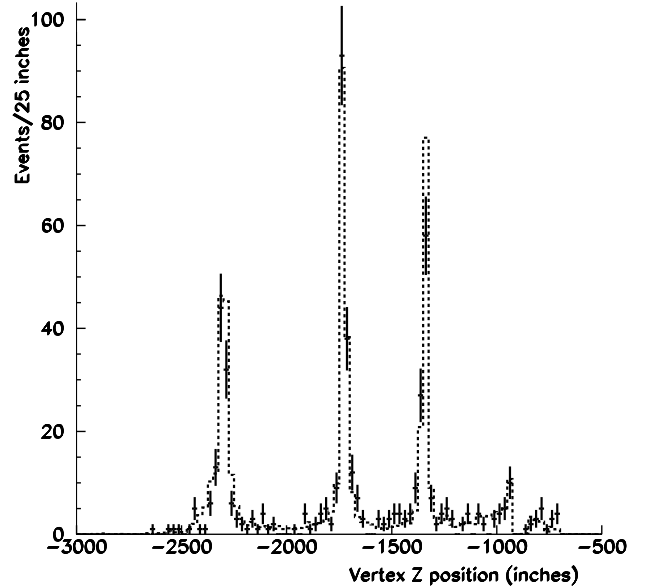,width=0.42\textwidth,bbllx=33pt,bblly=146pt,bburx=550pt,bbury=660pt,clip=T}}
\caption{ Longitudinal vertex position for all events in the decay channel
fiducial volume. (Crosses: data; histogram: Monte Carlo).  Peaks
correspond to interactions in veto wall and testbeam chamber (left) and 
drift chambers DK5 and DK4 (center and right).}
\label{hwmany}
\end{figure}

\vfill

\begin{figure}
\centerline{\psfig{figure=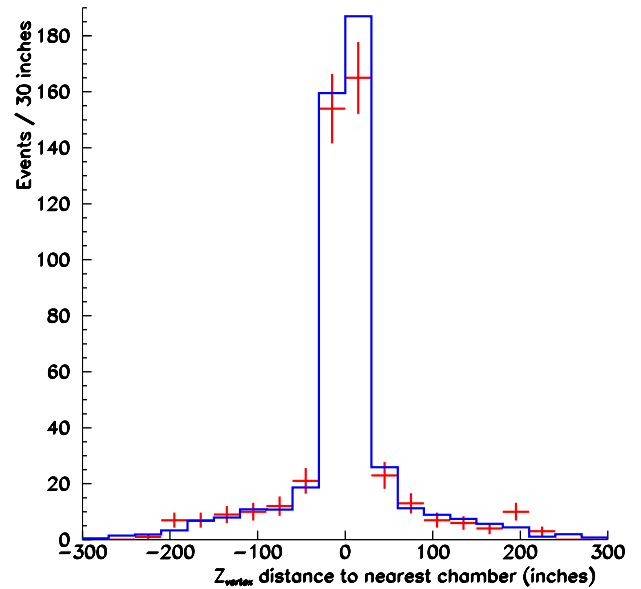,width=0.42\textwidth
,bbllx=33pt,bblly=146pt,bburx=550pt,bbury=660pt,clip=T}}
\caption{ Distance of longitudinal vertex position from the closest chamber for
all events. (Crosses: data; histogram: Monte Carlo.)}
\label{nearch}
\end{figure}

\vspace*{1.cm}

\newpage

\begin{figure}
\centerline{\psfig{figure=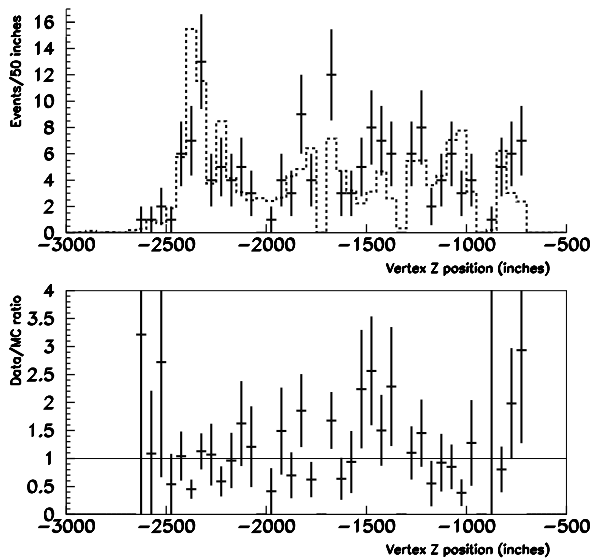,width=0.4\textwidth
,bbllx=33pt,bblly=150pt,bburx=550pt,bbury=660pt,clip=T}}
\caption{Top: longitudinal vertex position for events in the helium, where helium is defined as an event having a vertex more than 101.6 cm away from a chamber. (Crosses: data; histogram: Monte Carlo.) Bottom: ratio of data 
to Monte Carlo events.}
\label{heliumz}
\end{figure}

\noindent
The majority of
these events were from interactions in the chamber material 
or from interactions in the laboratory floor. In the data, 
from 502 events, 169 events had vertices reconstructed in the 
helium, defined as $>101.6$~cm for the nearest drift chamber.  This can be 
compared to Monte Carlo, which predicted $(525\pm 84)$ events 
with $(159\pm 25)$ events reconstructed 
in the helium.  Comparisons of data to Monte Carlo vertex 
distributions are shown in Figures~\ref{hwmany},~\ref{nearch}, 
and~\ref{heliumz}.

\begin{figure}
 \centerline{\psfig{figure=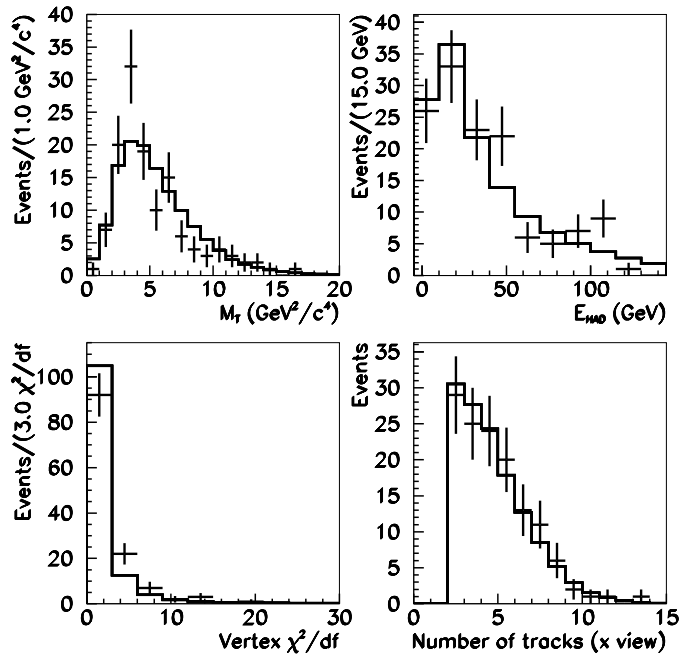,width=0.5\textwidth}}
 \caption{Distributions comparing data (points) and Monte Carlo (histogram)
  for charged-current DIS interactions in the decay channel chambers.  The
  distributions are: (a) transverse mass; (b) hadronic energy in the
front of the calorimeter; (c) vertex pseudo-$\chi^2$/dof; (d) number of
reconstructed tracks in the $x$ view. \label{fig:tordist}}
\end{figure}

The second sample was comprised of Monte Carlo events passing the 
five normalization cuts. Figure~\ref{fig:tordist} shows a comparison 
of some event variables for this sample.  There is good, 
qualitative agreement between the data and MC in these and 
additional distributions.  For events with a $z$ vertex more 
than 101.6 cm from the chambers (helium events) the Monte Carlo
predicts 28 events; 40 data events are observed.


\begin{table}
\caption{Estimated rates of background to the $N^0\rightarrow \mu \mu (\nu )$
search \label{backg}}
\begin{tabular}{lc}
Source & $\mu \mu (\nu )$ events \\ \hline
DIS events        & (3.9 $\pm $ 0.9) $\times $ 10$^{-2}$ \\ 
Diffractive charm & (1.1 $\pm $ 0.1) $\times $ 10$^{-3}$\\ 
Diffractive $\pi$ & (1.7 $\pm $ 0.1) $\times $ 10$^{-4}$ \\ 
Diffractive $K$   & (3.3 $\pm $ 0.3) $\times $ 10$^{-7}$ \\ 
K$_{L}^{0}$ decays from berm & (3.9 $\pm $ 3.9) $\times $ 10$^{-4}$ \\ 
Other sources & $ \ll$ 2.5 $\times$ 10$^{-4}$ \\
\hline
Total $\mu \mu (\nu )$ Background & (4.0 $\pm $ 0.9) $\times $ 10$^{-2}$ \\ 
\end{tabular}
\end{table}

\newpage

\bigskip

After all cuts, the preliminary expected background is 
0.040 $\pm$ 0.009 events in $\mu \mu$ mode, 
0.14 $\pm$ 0.02 events in $\mu e $ mode and 
0.13 $\pm$ 0.02 events in $\mu \pi$ mode.
As an example 
of the relative sizes of the contributions discussed above, 
the background sources for the $N^{0}\rightarrow \mu \mu (\nu )$ 
mode are broken down in Table~\ref{backg}.  

\section{CROSS-CHECKS USING DATA}

Before looking at the data in the signal region, we performed a series
of analyses on other fiducial and kinematic ranges which gave us
confidence in our Monte Carlo predictions.    We point out that the
two previous published analyses in the 0.3 to 3.0 GeV/$c^2$  and low mass region
were examinations of other such kinematic regimes.   Extra studies 
performed for this analysis included using 1) identical analysis 
cuts applied to events within $\pm$15.2 cm of a chamber 
(the chamber region); 2) the chamber region with loosened cuts to
increase $\mu \pi$ acceptance; 3) the ``intermediate region'' between
15.2 and 101.6 cm from the chambers, with otherwise standard analysis
cuts; and 4) events with well-reconstructed two-track vertices 
where the tracks were both identified as pions.  The results were 
within 1.5$\sigma$ of prediction in the above cases.

\begin{table}
\caption{Number of events in drift chambers which pass $N^0$ topology 
and ``clean'' cuts.\label{ab1}}
\begin{tabular}{ccc}
Decay Mode & Predicted Events & Observed Events \\ \hline
$\mu \mu$ (chamber) & 1.6 & 0 \\ 
$\mu e$ (chamber) & 1.8 & 1 \\ 
$\mu \pi$  (chamber) & 2.7 & 2 \\ 
\end{tabular}
\end{table}

\medskip

We use the chamber region events as an example of these studies.
This is a very powerful data sample because if any observed events 
in the decay region were due to neutrino-He interactions, then 
there should be a factor of 28 more events in the chambers, after 
scaling for acceptance and mass.  The numbers of observed events 
for the $\mu \mu ,$ $\mu e$, and $\mu \pi$ modes are listed in 
Table~\ref{ab1} along with the prediction for neutrino deep-inelastic 
scattering in the chambers. The observed events are consistent 
with the prediction, giving no indication of unexpected ``clean'', 
two-track neutrino interactions in the chambers. 
\vspace{-0.6cm}

\section{RESULTS OF THE SEARCH}
\vspace{-0.3cm}

Using the signal event selection criteria given above, the 
Monte Carlo background predictions are given in Table \ref{data_stats}. 
The number of observed events is also shown.  Three $\mu \mu(\nu )$ 
events were observed, which is considerably above the predicted background.  
No $\mu e$ or $\mu \pi$ events are observed, consistent with expectation.
\vspace{-0.5cm}

\begin{table}
\caption{Predicted and observed events passing all signal cuts
\label{data_stats}}
\begin{tabular}{ccc}
Decay Mode & Predicted Events & Observed Events \\ \hline
$\mu \mu (\nu )$ & $0.040\pm 0.009$ & 3 \\ 
$\mu e\left( \nu \right) $ & $0.14\pm 0.02$ & 0 \\ 
$\mu \pi $ & $0.13\pm 0.02$ & 0 \\ 
\end{tabular}
\end{table}


\begin{figure}
 \centerline{\psfig{figure=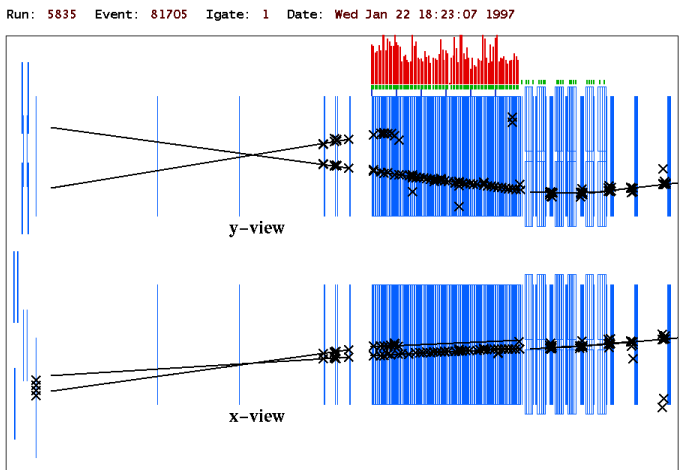,width=0.5\textwidth,clip=}}
 \caption{Run/Event 5835/81705: $\mu\mu (\nu)$ data event passing final cuts.}
 \label{evt3}
\end{figure}
\begin{figure}
 \centerline{\psfig{figure=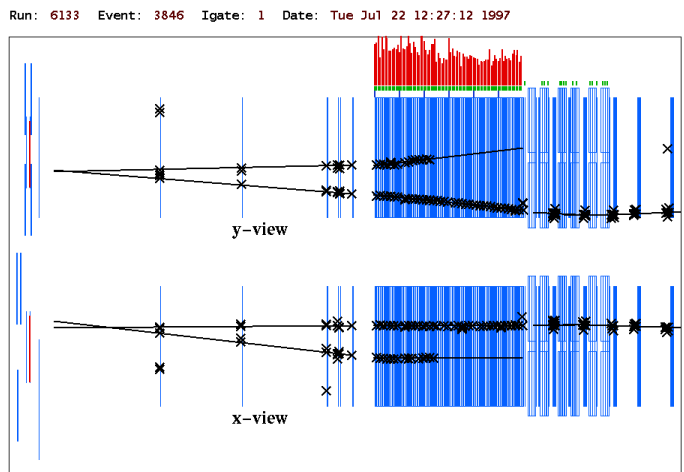,width=0.5\textwidth,clip=}}
 \caption{Run/Event 6133/3846: $\mu\mu (\nu)$ data event passing final cuts.}
 \label{evt2}
\end{figure}
\begin{figure}
 \centerline{\psfig{figure=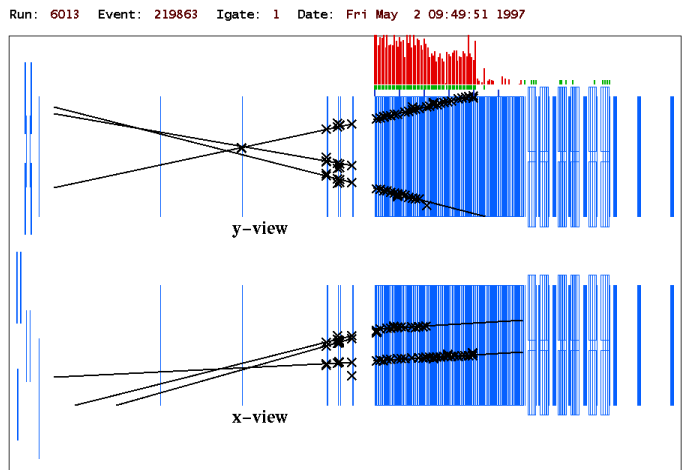,width=0.5\textwidth,clip=}}
 \caption{Run/Event 6013/219863: $\mu\mu (\nu)$ data event passing final cuts.}
 \label{evt1}
\end{figure}

The events are shown in Figs. \ref{evt3}-\ref{evt1}.   A summary 
of the event reconstruction characteristics is shown in 
Tables~\ref{evt_kine} and \ref{tab:vertex}. Because of the large 
multi-hit TDC ring-buffer used for the veto wall, it is typical for 
events to have an average of 1.7 counters firing per neutrino 
interaction.   For event 5835/81705, the TDC times with respect 
to the trigger are $+$404, $+$536 ns, where the positive sign
indicates the hits occurred after the trigger; for event
06133/03846, the hit occurred at $+24$ ns; and for event 6013/219863,
hits occurred at $-$256, $+320$, $+320$, and $+1192$ ns. Given 
the 3.8 ns timing resolution of the veto counters, these hits are 
well out of time.

\newpage

\begin{minipage}{\textwidth}
\begin{center}
\begin{minipage}{12cm}
\begin{table}
\caption{Kinematic and reconstruction quantities for the three 
candidate $N^0\rightarrow \mu\mu(\nu)$ events. The sign on the 
muon energy refers to the charge of the muon (if measured). \label{evt_kine}} 
\begin{tabular}{ccccccc}
Run/Event & Mode & E$_{\mu 1}$ & E$_{\mu 2}$ & $P_{T{\rm miss}}$
& $m_{\rm inv}$   & $m_{T}$  \\
          &      &  (GeV)      &   (GeV)     & (GeV/$c$)
& (GeV/$c^2$) & (GeV/$c^2$) \\ \hline
5835/81705 & $\nu $ & --77.7 & $\pm 2.56$ & 2.42 & 1.10 & 5.08 \\
6133/3846 & $\nu $ & --92.0 & $\pm 5.85$ & 1.41 & 0.88 & 3.08 \\
6013/219863 & $\nu $ & $\pm $48.0 & $\pm 4.34$ & 2.07 & 3.57 & 4.66 \\
\end{tabular}
\end{table}
\end{minipage}
\end{center}
\end{minipage}

\bigskip

\begin{minipage}{\textwidth}
\begin{center}
\begin{minipage}{12cm}
\begin{table}
 \caption{Vertex information for the observed $\mu\mu$ events; $\mathcal{P}_{\geq}=$ probability of an event having 
this pseudo-$\chi^{2}$/dof or greater. 
    \label{tab:vertex}} 
 \begin{tabular}{ccccc}
  Run/Event & $v_x$ & $v_y$ & $|z_{\rm vertex}-z_{\rm chamber}|$ &
   pseudo-$\chi _{\rm vertex}^{2}/dof$ \\ 
     & $(cm)$ & $(cm)$ & $(cm)$ & ($\mathcal{P}_{\geq}$)  \\ \hline
  5835/81705  & --46.7 & 3.6 &  196 & 6.3/9 (62\%) \\ 
  6133/3846   & 46.5 & --38.6 & 792 & 166/17 (5\%) \\
  6013/219863 & --59.2 & 14.7 & 186 & 22.6/10 (21\%)
 \end{tabular}
\end{table}
\end{minipage}
\end{center}
\end{minipage}

\section{CROSS-CHECKS BY RELEASING CUTS}

The observation of the three $\mu\mu$ data events prompted further
tests comparing data to Monte Carlo predictions with reduced cuts.
These studies provide cross-checks on whether the signal was
manufactured by the cuts, whether the Monte Carlo accurately models the
data just outside the cuts, and whether there is any indication of
excess background with less restrictive requirements.

The first study examined how data and Monte Carlo rates varied as the
cuts were loosened gradually. The expected background and number of
observed events at each step of the process is shown in Table
\ref{data_loose1}. The first step was to remove the ``clean cuts.'' At
this point, one additional $\mu \mu $ event was observed. The total is
written as $3+1$, explicitly separating off the three candidates. The
topology of this additional event was in agreement with the topology 
for a DIS event (extra track segments, excess of neutral energy, and 
extra hits), and with the Monte Carlo estimate of 0.25 events. 
No $\mu e$ or $\mu \pi $ events were observed which is consistent 
with expectation. Thus, except for the excess three events,
at this step there is agreement between data and Monte Carlo.
Continuing to remove the cuts sequentially, the fiducial region
was increased to include the chambers; finally, all energy and PID
cuts were removed. In both of these steps, the data continued to be in
agreement with the Monte Carlo, except for the three candidate events.

\newpage

\vspace*{11cm}

A second study considered the effect of loosening the cuts on the
expectation for a Monte Carlo which combined $N^{0}$ decays in $\mu \mu \nu$ 
mode with the standard background Monte Carlo. This study normalized
the $N^{0}$ decay Monte Carlo to 3 events for standard cuts and then
observed the expectation with the cuts sequentially released as shown
in Table~\ref{data_loose2}. There is agreement of the data with the
combined model but the low statistics are not definitive.

A third cross-check involved alternately releasing and then returning
individual cuts and cut-pairs. For most cuts, no extra events entered
the sample. When the requirement of ``no extra track segments'' was
removed, one $\mu \pi$ event entered the sample. When the ``no extra
clusters'' cut was removed, one event identified as $\mu e$ entered
the sample. This is consistent with the Monte Carlo expectation, where
the extra cluster is due to photons from a $\pi^0$ and these cause the
event to look electromagnetic. When both of these cuts are removed,
two more events enter the sample. Their topology is consistent with
DIS. In summary, this cross-check did not reveal events which are
similar to the candidates.

\clearpage

\vfill

\begin{minipage}{\textwidth}
\begin{center}
\begin{minipage}{15cm}
\begin{table}
\caption{Observed events and expected background as cuts are sequentially
released.\label{data_loose1}}
\begin{tabular}{c|c|c|c}
Sequential Change to Cuts & Event Type & Number of Data Events &  Monte Carlo Prediction\\
\hline
All cuts
& $\mu\mu$ & $3$ & 0.04 \\
& $\mu\pi$ &  0  & 0.14 \\
& $\mu e$  &  0  & 0.13 \\ 
\hline
Remove the ``Clean Cuts''
& $\mu\mu$ & $3+1$ & 0.25 \\
& $\mu\pi$ &   0   & 0.23 \\
& $\mu e$  &   0   & 0.70 \\ 
\hline
Include the chamber region
& $\mu\mu$ & $3+2$ & 1.3 \\
& $\mu\pi$ & 7     & 7.8 \\
& $\mu e$  & 5     & 4.7 \\ 
\hline
Release energy \& PID requirements
& $\mu\mu$ & $3+3$ &  2.3 \\
& $\mu\pi$ &  10   & 16.0 \\
& $\mu e$  &  10   & 10.8 \\ 
\end{tabular}
\end{table}
\end{minipage}
\end{center}
\end{minipage}

\vspace{2cm}

\begin{minipage}{\textwidth}
\begin{center}
\begin{minipage}{15cm}
\begin{table}
\caption{Predicted and observed $\mu\mu$ events as cuts are sequentially
released. Prediction shows the expected background and the combination of
signal (from  $N^0$ decay Monte Carlo) and background.   The signal is
normalized to 3 events with all cuts.   
\label{data_loose2}}
\begin{tabular}{c|c|c|c|c}
Event Type & Predicted Background & $N^0$-decay & $N^0$-decay + Background & Data \\
\hline
$\mu\mu$ with ``standard cuts''  &  0.04  &  3.00  &  3.04  & 3\\
releasing the ``clean cuts''     &  0.25  &  5.30  &  5.55  & 4\\
Including chamber regions        &  1.3   &  6.1   &  7.5   & 5\\
removing $E$ \& PID requirements &  2.3   &  6.1   &  8.6   & 6\\
\end{tabular}
\end{table}
\end{minipage}
\end{center}
\end{minipage}

\vfill

\clearpage

\section{CONSIDERATIONS UNDER AN $N^0$ HYPOTHESIS}

\begin{figure}
 \centerline{\psfig{figure=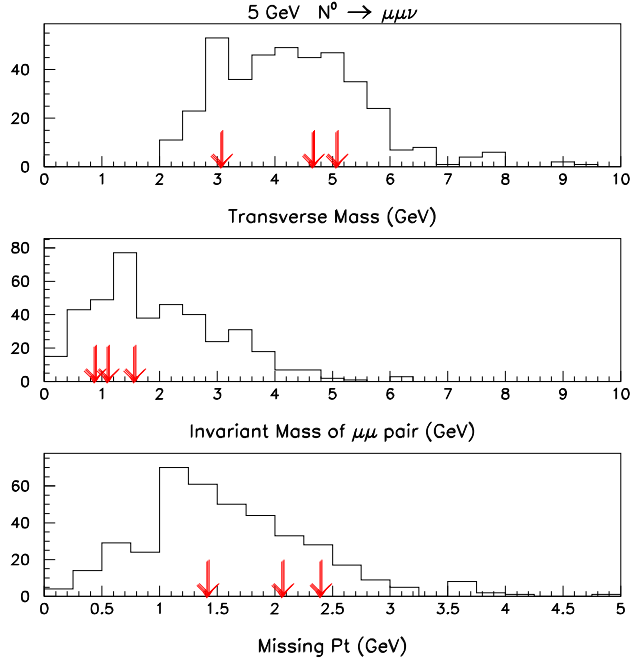,width=0.5\textwidth}}
 \caption{Kinematic distributions (transverse mass, invariant mass and
  missing transverse momentum) for the 5.0 GeV/$c^2$ $N^0$ Monte
  Carlo.  The histograms show the MC; the arrows indicate the
  three observed events.}
 \label{fig:n0dists}
\end{figure}

In many ways, the three $\mu\mu$ events are consistent with a $N^{0}$
decay hypothesis. The events pass the analysis cuts, where the background
is estimated to be 0.04 events. As expected for a decay relative to 
an interaction hypothesis, all three events occur well within the 
fiducial volume away from the chambers and are evenly distributed 
throughout the decay channel. The transverse mass, invariant mass, 
and missing $P_{T}$ are all consistent with a 5 GeV/c$^{2}$ $N^{0}$ 
decay (Fig.~\ref{fig:n0dists}).

Unlike the background, in both the NHL and neutralino models, one would
expect the $\mu \pi $ rate to be highly suppressed relative to leptonic
decays. However, for a 5 GeV/c$^{2}$ NHL model, one would expect 1.4
times more $\mu e\nu $ events~\cite{tim}. This is not inconsistent with the
observation of no $\mu e$ candidates, but neither does it provide direct 
support. A neutralino model, on the other hand, can accommodate the 
observation of either only $\mu \mu $ or a combination of $\mu\mu$ 
and $\mu e$ candidates by selecting appropriate couplings.

\begin{figure}
 \centerline{\psfig{figure=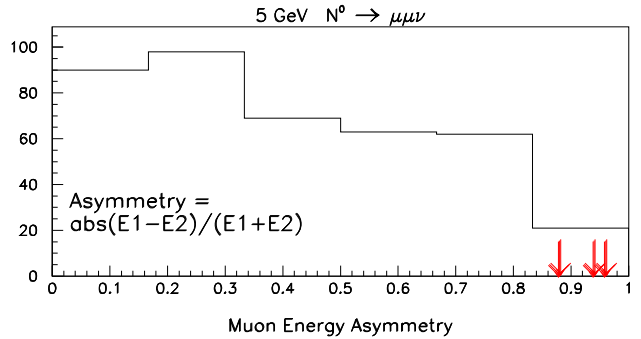,width=0.5\textwidth}}
 \caption{The muon energy asymmetry \mbox{($|E_{1}-E_{2}|$)/($E_{1}+E_{2}$).}
  The histogram shows the 5 GeV/$c^2$ $N^0$ MC;the arrows indicate
  the three observed events.}
 \label{n0asym}
\end{figure}

Globally, the events have one feature which is improbable for an
$N^{0}$ decay hypothesis. Figure~\ref{n0asym} shows the muon energy
asymmetry ($|E_{1}-E_{2}|$)/($E_{1}+E_{2}$) for the 5 GeV/c$^{2}$
$N^{0}$ simulation compared to the three data events. All three events
have a muon energy asymmetry which is greater than 0.85. The
probability that this occurs in a weak decay hypothesis~\cite{joe} 
is less than $0.5\%$ (including acceptance).

Individually, there are also particular characteristics to note in two
of the events. Event 06133/03846 has relatively poor vertex 
pseudo-$\chi^2$/dof; the probability of an event having a 
pseudo-$\chi^2$/dof greater than or equal to this is 5\%. In event 
6013/219863, three of the four hits in chamber DK4 immediately 
downstream of the vertex are missing.  The probability of this 
occurring in a random event is less than $3 \times 10^{-5}$.  
DK4 had persistent high voltage problems throughout the run and 
the absence of hits in this chamber may be due to this.  However, 
we have examined the drift chamber readout string and found no 
indication that the chamber was misbehaving for this event and 
we have observed good tracks through this chamber for events 
proximate in time.  Finally, we have discovered some evidence 
for a ``coherent'' inefficiency, a linked inefficiency between 
the $x$ and $y$ views.  Such an inefficiency may occur at a level 
of ($1.0 \pm 0.7$)$\times 10^{-3}$ probability, again making the 
missing of the three hits on the tracks highly unlikely.  This 
event also has a third track attached to a downstream vertex. 
To interpret this event as an $N^0$ decay, the added track 
must be interpreted as a delta ray.  Such an interpretation is 
consistent with the calorimeter data. Event 5835/81705 does not 
suffer from reconstruction issues.

\vspace{2cm}

\vfill

\vspace{2cm}

\newpage

\begin{figure}
\centerline{\rotate[l]{\psfig{figure=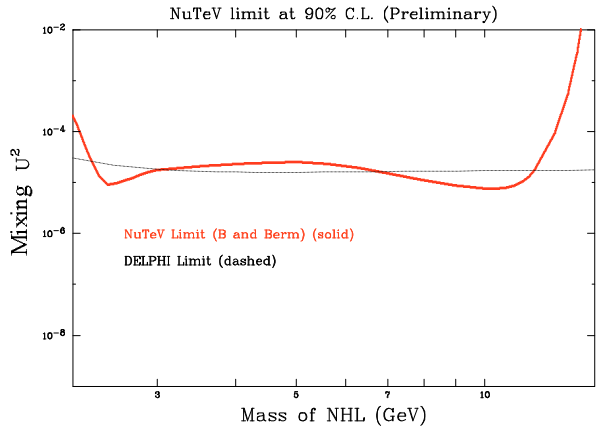,width=0.4\textwidth,clip=T}}}
 \caption{NuTeV limit on NHL production from B decays and berm production.}
 \label{nhllim}
\end{figure}

\begin{figure}
\centerline
{\psfig{figure=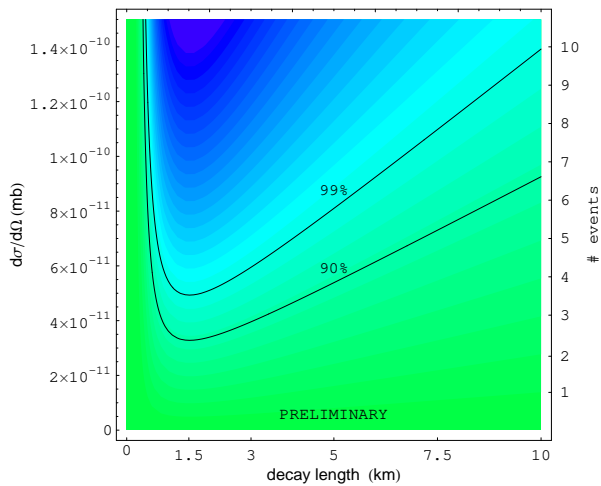,bbllx=60bp,bblly=30bp,bburx=470bp,bbury=340bp,width=0.5\textwidth,clip=T}}
 \caption{NuTeV limit on neutralino production. This limit is generic for
   an $N^0$ produced at the target.  The right axis (\# of events) follows
   the shaded contours.}
 \label{neutlim}
\end{figure}

Interesting limits on the production of NHLs and neutralinos can be
set that indicate the sensitivity of the decay channel search; 
these limits are set by calculating one-sided limits using a frequentist
approach without background subtraction. 
Because the source of the events is unclear, the unified approach 
of Feldman and Cousins~\cite{feldman} has deliberately {\em not} 
been used. The limit on NHLs reaches mixing parameter values 
below $|U|^2 = 10^{-5}$ and is consistent with the results from 
the Delphi experiment at LEP~\cite{DELPHI}, as shown on 
Figure~\ref{nhllim}. NuTeV is the first experiment to set limits on the
production of long-lived neutralinos in this mass range which decay by 
$R$-parity violation. The result is shown in Figure~\ref{neutlim}. This limit,
although motivated by a neutralino hypothesis, is a generic limit
applicable for any model of neutral particle production at the 
target~\cite{liubo_susy}.

\section{CONSIDERATIONS UNDER VARIOUS NULL HYPOTHESES}

\begin{figure}
 \centerline{\psfig{figure=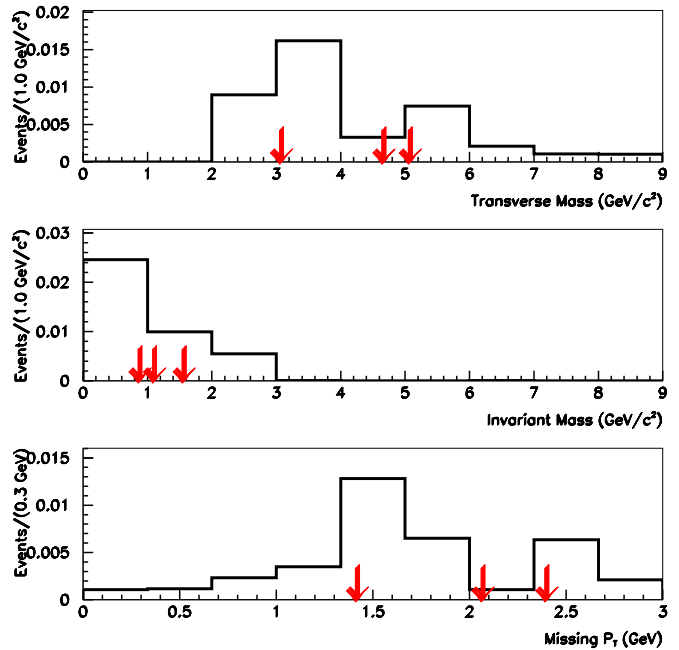,width=0.5\textwidth}}
 \caption{Kinematic distributions (transverse mass, invariant mass and
  missing transverse momentum) for $\mu\mu$ events from the background
  Monte Carlo.  The histograms shows the MC while the arrows indicate the
  three observed events.  \label{fig:disdists}}
\end{figure}

\begin{figure}
 \centerline{\psfig{figure=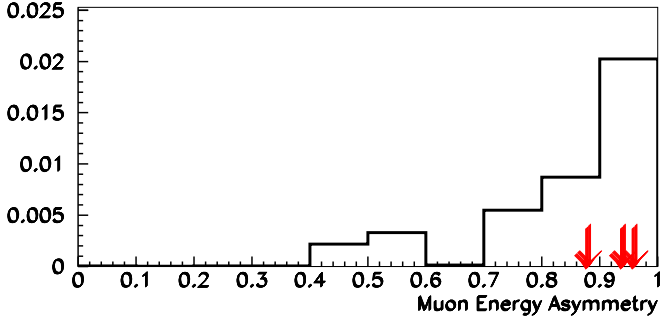,width=0.5\textwidth}}
 \caption{The muon energy asymmetry ($|E_{1}-E_{2}|$)/($E_{1}+E_{2}$).
  The histogram shows the $\mu\mu$ background Monte Carlo while the arrows 
  indicate the three observed events.  \label{fig:disasym}}
\end{figure}

Several aspects of the candidate events are similar to those from
neutrino interaction backgrounds, and might be indicative of 
unaccounted-for sources. First, all three events occurred during 
the high rate $\nu $-mode as opposed to $\overline{\nu} $-mode 
running periods; the $\nu $ to $\overline{\nu}$ event ratio is 
expected to be 4:1 for all events and 1.5:1 for low-multiplicity 
events. Second, if the events were produced by neutrino interactions,
then one would expect a high energy leading $\mu ^{-}$ for $\nu $-mode
running.  In the two cases where the charge of the leading muon can be
measured, the charge is negative.   The event kinematics ($M_T$,
$M_{\mu\mu}$, $P_T$) are consistent with the background simulation
(Fig.~\ref{fig:disdists}) and, for DIS background, the
probability for three events with the observed energy asymmetry is
25~to~$35\%$ (Fig.~\ref{fig:disasym}).   If these are interpreted as
DIS events, then the visible (measured) squared four-momentum transfer, 
$Q_{vis}^2$, for each is 6.4, 2.5, and 5.6 GeV$^2$, respectively.   

\bigskip

On the other hand, it is difficult to explain these events as neutrino
interactions.  If all of these events are due to charged-current neutrino
interactions in the helium which give a prompt second muon (charm
production, trident production, etc.), then the rate is two orders of
magnitude greater than predicted by the Monte Carlo.  If such a source was
not included in our simulations, then we would also expect between 10 and
300 additional events at 99\% confidence level~\cite{feldman} in the chambers 
and between 30,000 and 220,000 events at 99\% confidence level
in the calorimeter itself.  We actually see only 
2 $\mu\mu$ chamber events in agreement with our Monte Carlo
(see Table~\ref{data_loose1}) and 54 events with similar kinematics were 
found in the calorimeter~\cite{diffnutev}.  In order to show the difficulty
of developing reasonable ``null'' (background) hypotheses, we consider
three examples below.

The first null hypothesis is that the events were actually prompt $\mu \mu$
events produced in the chambers and misreconstructed in the helium.
If we try to force vertex reconstruction for the two muons in 
6013/219863 at the nearest downstream chamber (DK4), the 
pseudo-$\chi _{\rm vertex}^{2}/{\rm dof}$ goes from 22.9/10 for 
the standard fit up to 290.2/11 for the forced fit.   If the third 
track is also forced to be on the vertex, the 
pseudo-$\chi _{\rm vertex}^{2}/{\rm dof}$ rises to 323.6/17.
If one does not require the track to be in the chambers, but requires
all three tracks to come from the same vertex, then one obtains
217.1/16.   Likewise, event 6133/3846 produces a poor fit to the
chambers.   For the chamber downstream of the vertex (DK5), the 
pseudo-$\chi_{\rm vertex}^{2}/{\rm dof}$ is 94,720/18 and for 
the testbeam chamber upstream it is 543/18.   These are well above the
original fit of 166/17.   Finally, for 5835/81705, the standard fit
gives 6.3/9 which increases to 51.5/10 if forced to be
reconstructed in the chamber upstream (DK4).   Aside from the problems
of forcing a fit to the chambers, this hypothesis leaves unanswered 
the questions: 1) why are only $\mu \mu$ events misreconstructed, and
not $\mu \pi$ or $\mu e$? and 2) what is the source of these excess $\mu
\mu$'s?

A second null hypothesis rejects the idea that the $\mu \mu$
events are prompt, and instead attributes them to unsimulated 
$\nu + X \rightarrow \mu + \pi + Y$ interactions, where the pion 
decays to produce the second muon and the other particles in the 
reaction (signified by $Y$) are not seen.  This has the advantage 
of being more plausible than an unexpected source of prompt 
$\mu \mu$ events, but it suffers from several problems. First, if 
these are events which occur in the chambers, then one must 
explain why the $\mu \mu$ events are misreconstructed in the
helium. The transverse momentum associated with pion decay is very
small, so the muon track would generally point back to the original
chamber vertex.   For those events which do show an offset, 
there will be a much higher probability of occurring in the 
15.2 to 101.6 cm range than well beyond 101.6 cm, where these events 
are found.   Also, only 7\% of the pions will decay within the 
decay channel so many undecayed $\mu \pi$ events should be observed.

Quantitatively, the data in these other channels can 
be compared to different scalings of the observed three
$\mu\mu$ events. The DIS Monte Carlo predicts a ratio of $\mu\pi$
to $\mu\mu$ events of 3.25 for the sample with all the standard 
cuts. With this ratio, the probability of seeing 3 $\mu\mu$ 
events and, as observed, 0 $\mu\pi$ is 0.31\%.  Furthermore, 
one can scale these $\mu\mu$ events in the helium by the 
ratio of masses and acceptance for the chambers versus helium 
and compare to the observed $\mu\pi$ and $\mu\mu$ events seen 
in chambers with the loosest cuts. Given that ratio, the 
10 $\mu\pi$ (2 $\mu\mu$) events observed in the chambers are 
consistent at a confidence level of only $2.1 \times 10^{-5}$ 
($2.12 \times 10^{-5}$) with the hypothesis that the three
$\mu\mu$ events in the helium are due to an unaccounted-for $\mu\pi$
source. It is therefore extremely unlikely that an unsimulated
$\mu\pi$ source is responsible for the three $\mu\mu$ events in 
the helium.

A third null hypothesis is that the candidates are neutrino
interactions in the chambers producing a $\mu K$ final state followed
by a $K \rightarrow \mu \nu$ decay that causes the event vertex to be
misreconstructed downstream in the helium region. This process has the
advantage that the decay angle is sufficiently large for an event
produced in the chambers to reconstruct in the helium. Fits to the
three candidate events under this hypothesis show that that the fits
are kinematically possible. Taking the kaon momenta calculated from
the fit, the probabilities are 17.6\%, 11.6\% and 8.2\% for the kaon 
to decay. Because the decay channel tracking is largely located 
downstream of the decay region, most two-track vertices, including 
most simulated $N^0$ events, can also be reconstructed as 
displaced-vertex kaon decays with high quality.

However, the kaon hypothesis does not explain these events for several
reasons. First, direct production is Cabibbo-suppressed, so most
production is through fragmentation.  Direct production is also most likely
in $\bar \nu$ running, whereas the events are only seen in $\nu$ running
periods. Approximately 20\% of events produce kaons through
fragmentation, but this occurs mainly in high multiplicity events which
would be cut by the ``clean cuts.'' This process is included in the
0.040 event background estimate, to which it contributes $\sim$0.0025
events.  Clean events could be produced by diffractive production;
the large angle of the high energy muons in these events, however, is 
uncharacteristic of diffractive production and the excess is far above 
expectation (see Table \ref{backg}).

Finally, as with the pion decay hypothesis only a small fraction, 22\%,
of the kaons from neutrino interactions will decay. In addition,
because kaon decays have large $P_T$, only 45\% of the $\mu-K$
producing two muons in the calorimeter will have a vertex formed by the
two muons. In the other cases, the decay plane is aligned such that the
two muons fail to verticize in the decay channel. (Such unaligned
dimuon events consistent with all but the vertex cut are not observed
in the data.) As in the case of the pions, we can use scaling
arguments to compare the observed data to the predictions of
this hypothesis.  In the data, there are one $\mu e$ and two $\mu\pi$
chamber events with all the standard cuts but with the track energy
requirements removed.  The scaling factor is 8.24 between $\mu K$ events 
in the drift chambers and $\mu\mu$ events in the helium.  Under this
hypothesis, the probability to see, as observed,  three or fewer 
$\mu$-hadron events in the chambers given the three $\mu\mu$ events in 
the helium is 0.37\%.  Therefore, the paucity of $\mu$-hadron events in 
the chambers excludes at 99.6\% confidence level the possibility
of an unsimulated $\mu K$ source large enough to explain the three 
observed $\mu\mu$ events in the helium.

\section{SUMMARY AND CONCLUSIONS}

In summary, NuTeV has observed 3 $\mu \mu$ events, 0 $\mu \pi$ and 0
$\mu e$ events with transverse mass above 2.2 GeV/$c^2$.   The expected
backgrounds were $0.040 \pm 0.009$,  $0.14 \pm 0.02$, and $0.13 \pm
0.02$ events (preliminary), respectively.  

NuTeV has set new limits on NHL and neutralino decays based on
this analysis.  The NHL limits are consistent with the Delphi result.
The neutralino results are the first in this kinematic region for the
long-lived $\chi^0$ which decays with $R$-parity violation.

In conclusion, the rate corresponding to the observed three events is not
consistent with Standard Model processes we have identified 
and the source of the events is not clear.

\bigskip

This research was supported by the U.S. Department of Energy and the
National Science Foundation. We thank the staff of FNAL for their
contributions to the construction and support of this experiment during the
1996-97 fixed target run.

\end{document}